\documentclass{ws-procs9x6}

\begin{document}

\title{Unconventional AGN from the SDSS}

\author{P.~B. HALL, G.~R. KNAPP, G.~T. RICHARDS, M.~A. STRAUSS}
\address{Princeton University Observatory, Princeton, NJ 08544}

\author{S.~F. ANDERSON}
\address{Astronomy Department, University of Washington, Seattle, WA 98195}

\author{D.~P. SCHNEIDER, D.~A. VANDEN BERK}
\address{Dept. Astronomy \& Astrophysics, 
Penn State Univ., University Park, PA 16802}

\author{D.~G. YORK}
\address{Astronomy~Dept.~\&~Enrico~Fermi~Institute,~Univ.~of~Chicago,~Chicago,~IL~60637}

\author{K.~S.~J. ANDERSON, J. BRINKMANN, S.~A. SNEDDEN}
\address{Apache Point Observatory, P.O. Box 59, Sunspot, NM 88349}


\maketitle
\abstracts{We discuss some of the most unusual active galactic nuclei (AGN)
discovered to date by the Sloan Digital Sky Survey (SDSS): the first broad
absorption line quasar seen to exhibit He\,{\sc ii} absorption, 
several quasars with extremely strong, narrow UV Fe\,{\sc ii} emission,
and an AGN with an unexplained and very strange continuum shape.
}

\section{Introduction}

The Sloan Digital Sky Survey (York et al. 2000; Fukugita et al. 1996;
Gunn et al. 1998; Hogg et al. 2001; Stoughton et al. 2002;
Smith et al. 2002; Pier et al. 2003) is obtaining optical spectra
for $\sim$10$^5$ quasars over $\sim$$\frac{1}{4}$ of the entire sky.
Through careful target selection (Richards et al. 2002) and sheer size, 
the SDSS includes numerous AGN with unconventional properties.
The high-quality, moderate-resolution SDSS spectra 
can be used to set the stage for the detailed multiwavelength studies
often needed to understand interesting quasar subclasses.
We illustrate this fact using several unusual AGN 
included in the SDSS Second Data Release (Abazajian et al. 2004).

\section{The first He\,{\sc II} $\lambda$1640 broad absorption line quasar}

\begin{figure}[b]
\centerline{\epsfxsize=4.29in\epsfbox{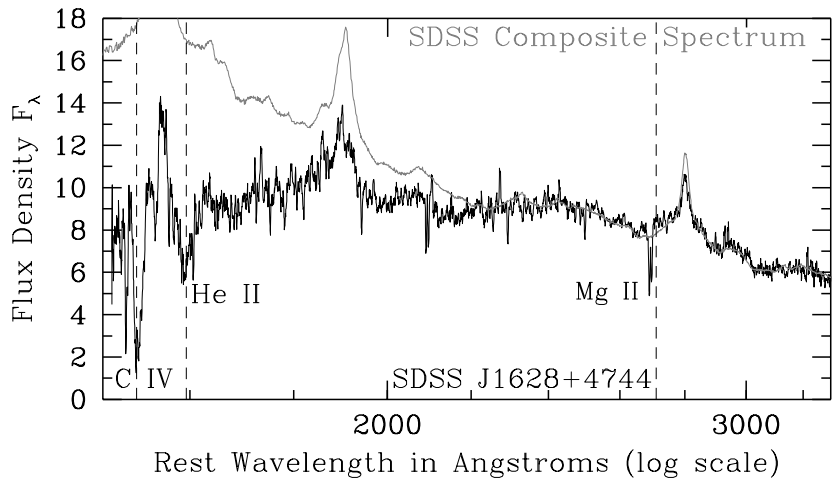}}
\caption{
The first He\,{\sc II} BAL quasar, SDSS J1628+4744.
The ordinate in this and all subsequent figures is flux density
$F_{\lambda}$ in units of 10$^{-17}$ ergs cm$^{-2}$ s$^{-1}$ \AA$^{-1}$.
\label{he2}} \end{figure}


Although broad absorption line (BAL) quasar outflows are known to include highly
ionized gas (e.g., Telfer et al. 1998), He\,{\sc ii} absorption had not been
detected in them until this meeting (see Maiolino et al. 2004 
for the second case).  
Figure\,\ref{he2} shows SDSS J162805.81+474415.6, 
a $z$=1.597 quasar with an outflow at $v$=8000 km s$^{-1}$ seen in C\,{\sc iv}
and He\,{\sc ii} $\lambda$1640.
He\,{\sc ii} $\lambda$1640 is analogous to H\,{\sc i} $\lambda$6563
(H$\alpha$) since it is the $n$=2$\leftrightarrow$3 transition.
The agreement between the velocity profiles of the C\,{\sc iv} and the putative
He\,{\sc ii} absorption is not exact, but is within the range of variation seen
between troughs of different ions in BAL quasars (e.g., Arav et al. 2001).  
There is also a narrow system at $z$=1.4967 ($v$=11800 km s$^{-1}$) seen in
C\,{\sc iv}, Al\,{\sc ii}, Fe\,{\sc ii} and Mg\,{\sc ii}
and an intervening, narrow Fe\,{\sc ii}+Mg\,{\sc ii} system at $z=0.9402$.



There 
are two other 
possible explanations for this trough.
First, it 
could be due to Al\,{\sc iii} at $v$=46000 km s$^{-1}$; 
the lack of accompanying Mg\,{\sc ii} absorption
would not be unprecedented (Hall et al. 2002).  
Detection of C\,{\sc iv}
absorption at 3450\,\AA\ would confirm this hypothesis.
Second, two SDSS quasars have Mg\,{\sc ii} absorption extending $\simeq$1500
km\,s$^{-1}$ redward of the systemic redshift (Hall et al. 2002); by analogy,
this trough could be C\,{\sc iv} redshifted by $\simeq$7000 km\,s$^{-1}$
(without accompanying Mg\,{\sc ii}).
That is an implausibly large velocity in terms of the 
redshifted absorption models discussed in Hall et al. (2002), but definitively
ruling out this possibility requires spectroscopy at $<$3800\,\AA\ to search
for redshifted Si\,{\sc iv} and N\,{\sc v}.

BAL quasars are thought to have large columns of highly ionized gas which
absorb X-ray but not UV photons (e.g., Chartas et al. 2002).  If the absorbing
gas is modeled as a slab whose illuminated face has ionization parameter $U$
(photon to H\,{\sc i}+H\,{\sc ii} density ratio), then the front of the slab is
a He\,{\sc iii} zone of equivalent column $N_H$$\simeq$$10^{21.8} U$,
followed by a He\,{\sc ii} zone of column $N_H$$\simeq$$10^{22.7} U$ and 
a He\,{\sc i}+H\,{\sc ii} zone of column $N_H$$\simeq$$10^{23} U$ 
(Wampler et al. 1995).
In this object we measure $N_{\rm HeII,n=2}$$\geq$10$^{15}$ cm$^{-2}$ 
(the lower limit applies if the trough is saturated).  
In the He\,{\sc ii} zone, $N_{\rm HeII}$$\simeq$0.1$N_H$, so as few
as 1 in 10$^{6.7}$$U$ He\,{\sc ii} ions in the $n$=2 state would
explain the observed He\,{\sc ii} $\lambda$1640 absorption.
But normal BAL quasar outflows do not show such absorption,
and so must have an even smaller He\,{\sc ii} $n$=2 population.
The two candidate He\,{\sc ii} BAL quasars known could differ from the norm
either by having extremely high ionization parameters $U$$\gg$10 (from gas 
at exceptionally small distances) or, more probably, high densities
$n_e$$\gg$10$^{10}$ throughout the He\,{\sc ii} region, such that collisional 
excitation of the $n$=2 state is non-negligible (Wampler et al. 1995).
Densities of at least $10^{10.5}$ cm$^{-3}$
are known to exist in BAL outflows (Hall \& Hutsem\'ekers 2003).

A full understanding of He\,{\sc ii} BALs will require photoionization modeling,
preferably in conjunction with wider wavelength coverage spectroscopy.

\section{Quasars with very strong and narrow UV Fe\,{\sc II} emission}


\begin{figure}
\centerline{\epsfxsize=4.5in\epsfbox{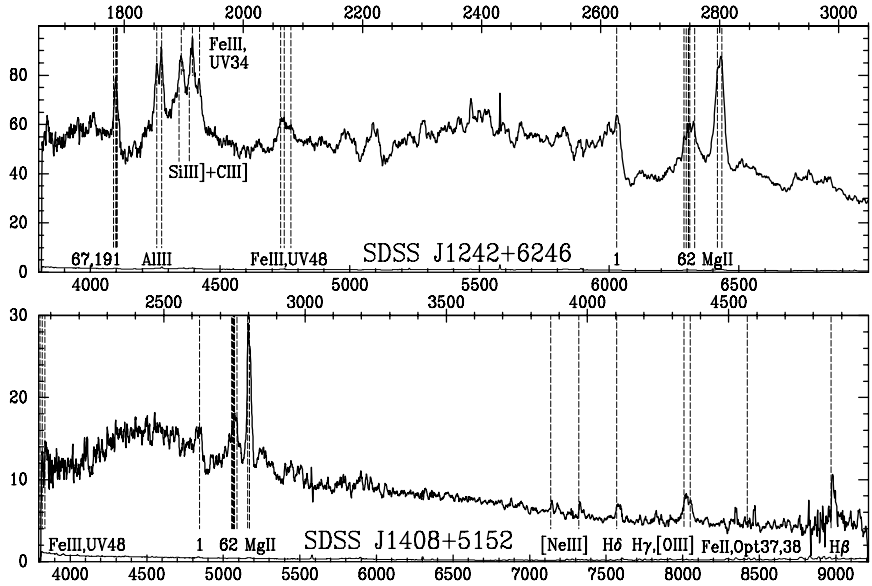}}
\caption{SDSS spectra of quasars with extremely strong, narrow UV
Fe\,{\sc ii} emission: 
a) SDSS J124244.37+624659.2; 
b) SDSS J140851.67+515217.4. 
Numbers indicate the ultraviolet Fe\,{\sc ii} multiplets
responsible for the emission at the wavelengths plotted.
\label{uvfe2}} \end{figure}

Fe\,{\sc ii} emission is very important to the energy balance of AGN broad
emission line regions (BELRs), but theoretical models have difficulty
reproducing the strength and shape of the observed Fe\,{\sc ii} complexes.
Bright objects with strong, narrow Fe\,{\sc ii} are thus extremely useful for
refining models and defining the areas of parameter space occupied by
BELRs.  Figure \ref{uvfe2} shows two such objects from the SDSS.
The very weak optical Fe\,{\sc ii} emission in SDSS J1408+5152 confirms that
the UV and optical emission strengths of Fe\,{\sc ii} can be highly
anticorrelated in individual objects (Shang et al. 2003).

\begin{figure}
\centerline{\epsfxsize=4.5in\epsfbox{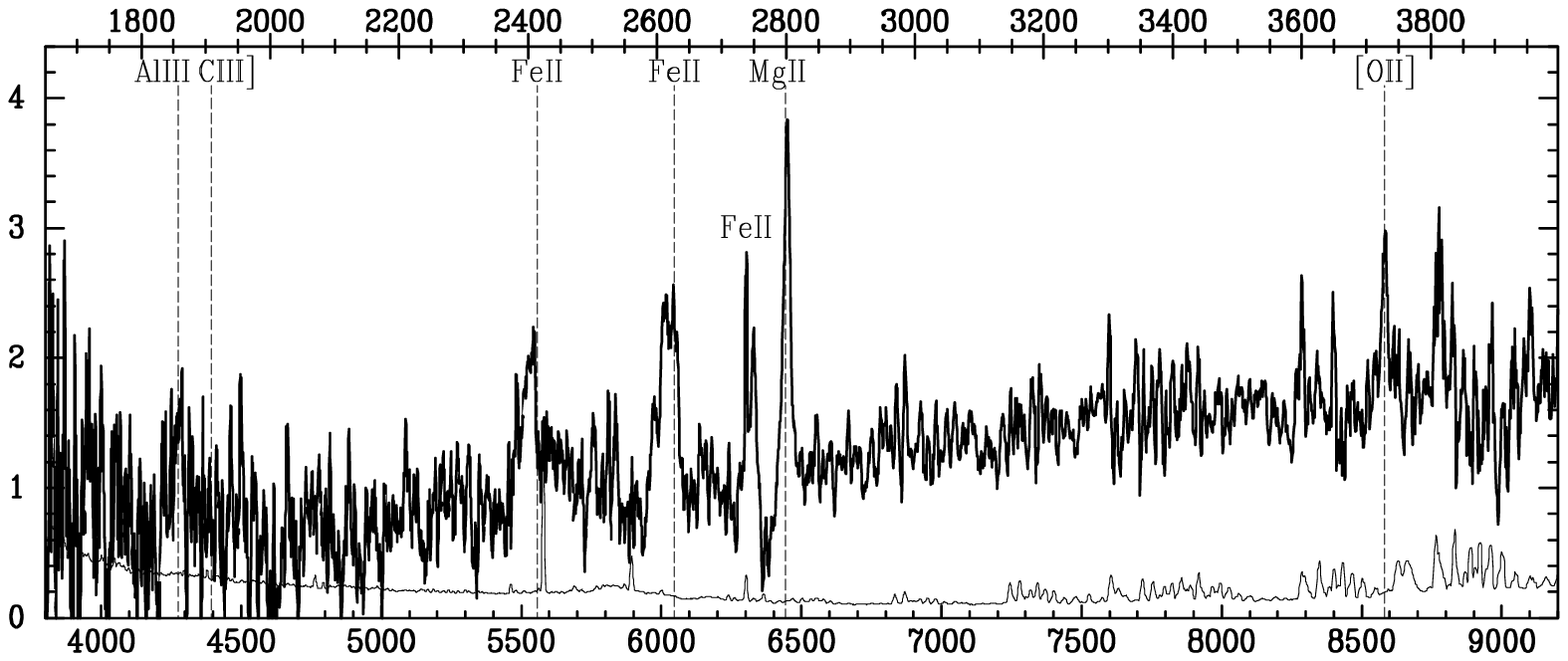}}
\smallskip
\centerline{\epsfxsize=4.5in\epsfbox{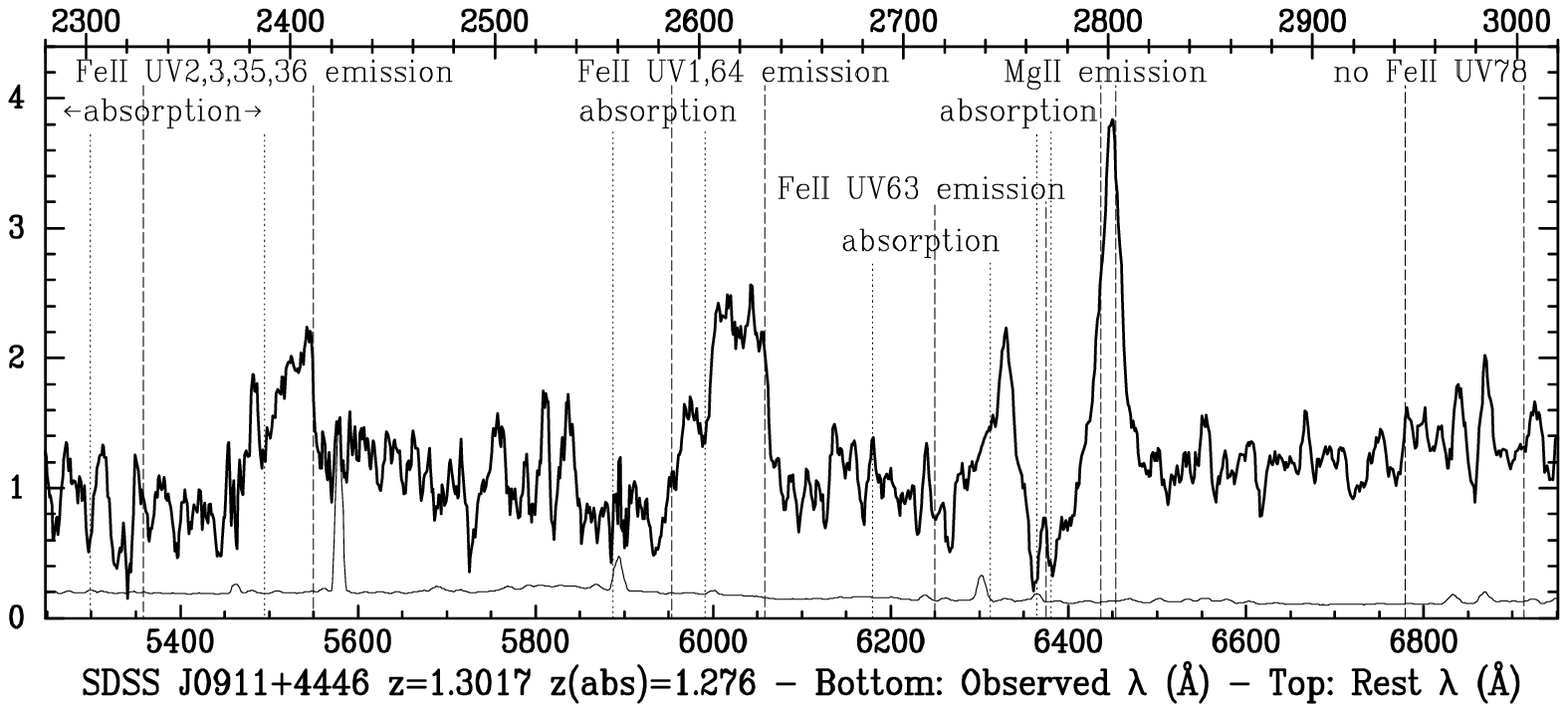}}
\smallskip
\centerline{\epsfxsize=3.5in\epsfbox{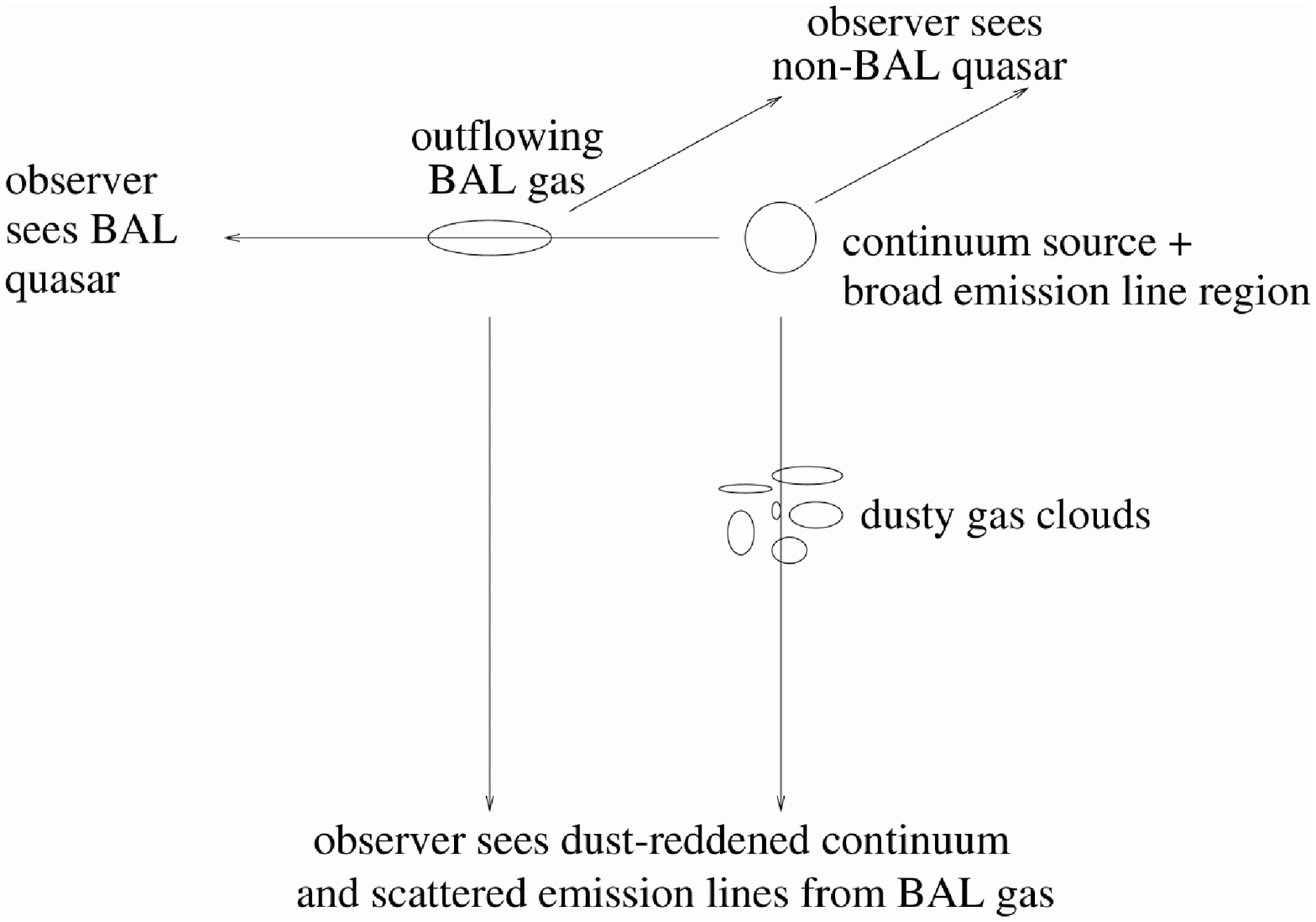}}
\caption{Top: the unusual Fe\,{\sc ii} emitter SDSS J0911+4446.
Middle: closeup of its spectrum shows Fe\,{\sc ii} multiplet emission at 
$z$=1.3017 (between dashed lines) and absorption at $z$=1.276 (between dotted
lines).
Bottom: a possible explanation wherein a strongly reddened continuum allows
Fe\,{\sc ii} emission scattered from a BAL outflow to be detected.
\label{fe2freak}} \end{figure}

Even more interesting is SDSS J091103.49+444630.4 
(Figure \ref{fe2freak}).  It shows Fe\,{\sc ii} emission and self-absorption
only in transitions involving lower energy levels $\leq$1~eV above ground
(UV78 at $\lambda$$\simeq$3000\,\AA\ has its lower level at 1.7\,eV, 
and is at best very weak).  The spectrum 
can be explained as a reddened continuum plus indirect (scattered) light
from a low-temperature BAL outflow 
(Figure \ref{fe2freak}c). 
Normally, such scattered emission is swamped by the direct spectrum, but 
it is entirely plausible that the direct spectrum could be obscured in 1 out of
10,000 quasars.
Alternatively, it may be a quasar where the Fe\,{\sc ii} emission is powered
only by photoionization, and thus closely matches theoretical expectations
(Baldwin et al. 2004).  
Improved 
spectra are needed to discriminate between these hypotheses.

\section{A REAL mystery object}

\begin{figure}
\centerline{\epsfxsize=4.5in\epsfbox{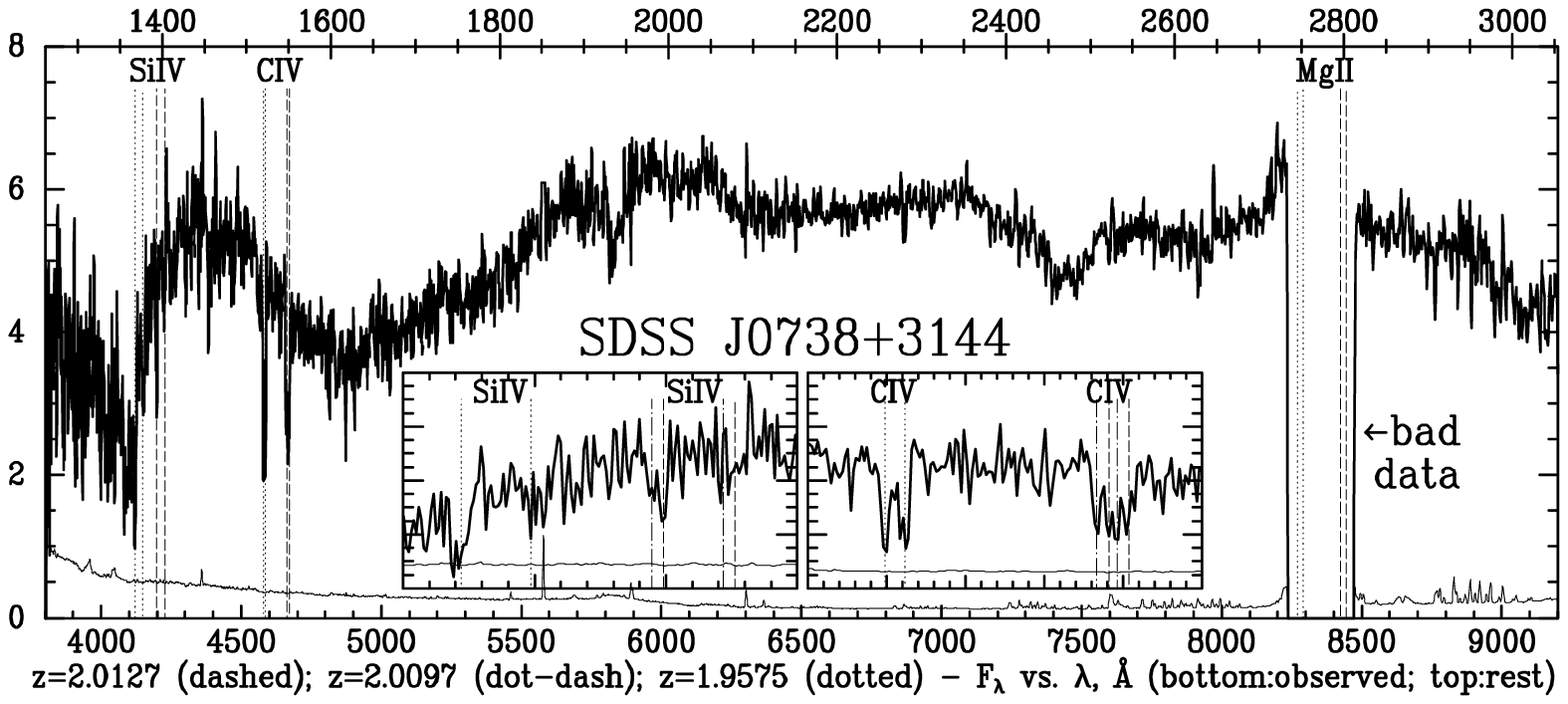}}
\caption{SDSS spectrum of SDSS J0738+3144, smoothed by a 3-pixel-wide boxcar.
The insets show regions of Si\,{\sc iv} and C\,{\sc iv} absorption near the
presumed redshift of $z$$\simeq$2.
\label{ucs}} \end{figure}

Lastly, we present SDSS J073816.91+314437.2 (Figure \ref{ucs}).  This object
is optically unresolved and was targeted only as a faint radio source.
Its redshift is probably $z$=2.0127, from narrow C\,{\sc iv} and Si\,{\sc iv} absorption,
with similar absorption systems at $z$=2.0097 and $z$=1.9575 (insets),  
and must be $z$$\leq$2.4, from the observed lack of Ly$\alpha$
forest absorption.  
But even though we know its redshift, we have no clear understanding of
the spectrum of SDSS J0738+3144.
Our best guess is that the spectrum contains broad, blueshifted emission in
Mg\,{\sc ii}, Fe\,{\sc iii} $\lambda\lambda$2080 (UV48), 
C\,{\sc iii}]+Fe\,{\sc iii} 
$\lambda\lambda$1915 (UV34), and possibly C\,{\sc iv}, plus BAL troughs of
C\,{\sc iv}, Al\,{\sc iii} and Mg\,{\sc ii} outflowing at 37,000 km\,s$^{-1}$
to explain the dips observed at 4100, 4900 and 7400\,\AA.
UV and NIR spectroscopy 
are needed to determine if idea is correct,
but the universe clearly 
contains some very unconventional AGN!

\section*{Acknowledgments}
\footnotesize{Funding for the SDSS (http://www.sdss.org/) has been provided by
the Alfred P.  Sloan Foundation, NASA, the NSF, the U.S. DOE, the Japanese
Monbukagakusho, the Max Planck Society \& the Participating Institutions  
(for whom the Astrophysical Research Consortium manages the SDSS):
U. Chicago, Fermilab, Institute for Advanced Study, Japan Participation Group,
Johns Hopkins U., Los Alamos National Laboratory, Max-Planck-Institute for
Astronomy, Max-Planck-Institute for Astrophysics, New Mexico State U., U.
Pittsburgh, Princeton U., U.S. Naval Observatory \& U. Washington.}


\begin{thebibliography}{0}

\bibitem{dr2} K. Abazajian {\it et al.}, 2004, {\it AJ}, submitted 
(astro-ph/0403325)

\bibitem{arav} N. Arav {\it et al.}, 2001, {\it ApJ}, 561, 118

\bibitem{bf} J. Baldwin {\it et al.}, 2004, to appear in {\it AGN Physics with
the Sloan Digital Sky Survey} (ASP: San Francisco), ed. G. Richards \& P. Hall

\bibitem{ab} M. Fukugita {\it et al.}, 1996, {\it AJ}, 111, 1748

\bibitem{gunn} J. Gunn {\it et al.}, 1998, {\it AJ}, 116, 3040

\bibitem{sb1} P. Hall {\it et al.}, 2002, {\it ApJS}, 141, 267

\bibitem{fe3} P. Hall \& D. Hutsem\'ekers 2003, in 
{\it Active Galactic Nuclei from Central Engine to Host Galaxy}
(ASP: San Francisco), ed. S. Collin {\it et al.}, 209

\bibitem{hogg} D. Hogg {\it et al.}, 2001, {\it AJ}, 122, 2129

\bibitem{heii} R. Maiolino {\it et al.}, 2004, {\it A\&A}, in press
(astro-ph/0312402)

\bibitem{pier} J. Pier {\it et al.}, 2003, {\it AJ}, 125, 1559

\bibitem{sdssqtarget} G. Richards {\it et al.}, 2002, {\it AJ}, 123, 2945

\bibitem{shang} Z. Shang {\it et al.}, 2003, {\it ApJ}, 586, 52

\bibitem{smith} J. Smith {\it et al.}, 2002, {\it AJ}, 123, 2121

\bibitem{edr} C. Stoughton {\it et al.}, 2002, {\it AJ}, 123, 485

\bibitem{wcp} E. Wampler, N. Chugai \& P. Petitjean, 1995, {\it ApJ}, 443, 586

\bibitem{sdss} D. York {\it et al.}, 2000, {\it AJ}, 120, 1579

\end{thebibliography}
\end{document}